\documentclass[a4paper]{article}

\usepackage{INTERSPEECH2022}
\usepackage{amsmath,graphicx,bm}
\usepackage[bookmarks=false]{hyperref}
\usepackage{algorithm}
\usepackage{algpseudocode}
\usepackage{booktabs}

\def\x{{\mathbf x}}

\title{EdiTTS: Score-based Editing for Controllable Text-to-Speech}
\name{Jaesung Tae$^{1*}$, Hyeongju Kim$^{2*}$, Taesu Kim$^3$}
\address{
  $^1$Yale University\\
  $^2$Supertone, Inc.\\
  $^3$Neosapience, Inc.}
\email{jake.tae@yale.edu, hyeongju@supertone.ai, taesu@neosapience.com}

\begin{document}

\maketitle

\begin{NoHyper}
\def\thefootnote{*}\footnotetext{Equal contribution. Work performed at Neosapience.}
\def\thefootnote{\arabic{footnote}}
\end{NoHyper}

\begin{abstract}
We present EdiTTS, an off-the-shelf speech editing methodology based on score-based generative modeling for text-to-speech synthesis. EdiTTS allows for targeted, granular editing of audio, both in terms of content and pitch, without the need for any additional training, task-specific optimization, or architectural modifications to the score-based model backbone. Specifically, we apply coarse yet deliberate perturbations in the Gaussian prior space to induce desired behavior from the diffusion model while applying masks and softening kernels to ensure that iterative edits are applied only to the target region. Through listening tests and speech-to-text back transcription, we show that EdiTTS outperforms existing baselines and produces robust samples that satisfy user-imposed requirements.\footnote{Code and audio samples at \url{https://editts.github.io}.}
\end{abstract}

\section{Introduction}

\label{sec:intro}
Score-based generative models~\cite{song2019generative} and denoising diffusion probabilistic models~\cite{ho2020ddpm} have proven remarkably successful in various generative tasks, achieving state-of-the-art performance in image generation~\cite{dhariwal2021diffusion,song2021scorebased} and speech synthesis~\cite{chen2021wavegrad,kong2021diffwave,jeong2021difftts,popov2021gradtts,popov2021diffusionbased}. This emerging class of generative models follows an iterative training scheme in which a complex data distribution is gradually corrupted with Gaussian noise, and the model learns to estimate gradient fields to reverse noisy priors back into their original distribution. Recent breakthroughs in image editing~\cite{meng2021sdedit,choi2021ilvr} also utilize score-based generative models to synthesize photo-realistic images from user edits, such as rough sketches or scribbles. In this setup, the pretrained model is effectively prompted to interpolate realistic details from coarse user input, producing an edited image.

Bearing strong resemblance to image editing, speech editing is a conditional generation task in which a model modifies some selected audio segments according to user instructions. Despite advancements in neural text-to-speech (TTS), fine-grained generation remains a non-trivial task, and it is often difficult to obtain a sample that exactly satisfies the user’s needs, especially when it comes to prosody and tone. This limitation highlights the need for a robust speech editing framework that can provide greater control over the final output. Previous works have used unit selection~\cite{jin2017voco} or context-aware prosody correction~\cite{morrison2021context}, but their reliance on traditional algorithms such as concatenation or TD-PSLA~\cite{moulines1989pitch} complicates the synthesis pipeline while compromising output quality. Alternative deep learning-based methods typically require training an additional editing model, such as a bidirectional fusion module~\cite{tan2021editspeech} on top of an existing TTS pipeline, incurring time and computation costs.

We propose EdiTTS, an off-the-shelf score-based speech editing framework that leverages a pretrained score-based backbone to enable granular editing of pitch and audio content. Unlike score-based image editing methods, which directly modify the input with user edits, EdiTTS induces editing behavior by applying coarse perturbations to diffused Gaussian priors. The score estimator then denoises the perturbed latents along the trajectory that best aligns with the data distribution to generate an edited sample. 

Our contribution can be summarized as follows: (1) We present a new method of editing speech, EdiTTS, which exploits the score-based TTS model to refine a coarsely modified mel-spectrogram prior. This conception allows for a unified approach to speech editing, where a single model is used to perform various editing tasks. (2) We show that EdiTTS is able to produce more natural, human-sounding speech without any conditional inputs such as pitch, thanks to the robust denoising process of the score-based TTS model. (3) EdiTTS is easy to implement since it leverages the existing TTS pipeline as-is without requiring any additional auxiliary module, fine-tuning, or changes to training objectives. In this work, we demonstrate our method using Grad-TTS~\cite{popov2021gradtts} as the model backbone, focusing on pitch shift and content replacement as popular applications of speech editing.

\section{Background}
\label{sec:format}

\subsection{Score-based Generative Modeling}

In score-based generative models, the original data distribution is corrupted with Gaussian noise according to an Itô SDE

\begin{equation} \label{eq:1}
    d \mathbf{x} = \mathbf{f}(\mathbf{x}, t) \, dt + \mathbf{g}(\mathbf{x}, t) \, d \mathbf{w},
\end{equation}
where $\mathbf{w}$ is the standard Wiener process, $t$ is a value sampled from a finite time interval $[0, T]$, and $\mathbf{f}(\cdot, t)$ and $\mathbf{g}(\cdot, t)$ are drift and diffusion coefficients of $\mathbf{x}$, respectively. 

A property of diffusion is that its reverse is also a diffusion process. Let $\bar{w}$ denote the reverse-time Wiener process. The reverse-time SDE~\cite{anderson1982reverse} is given by

\begin{equation} \label{eq:2}
\begin{split}
d \mathbf{x} 
&= [\mathbf{f}(\mathbf{x}, t) - \mathbf{g}(\mathbf{x}, t) \mathbf{g}(\mathbf{x}, t)^\top \nabla_\mathbf{x} \log p_t (\mathbf{x})] \, dt \\ 
&+ \mathbf{g}(\mathbf{x}, t) \, d \mathbf{\bar{w}}.
\end{split}
\end{equation}

Furthermore, Kolmogorov's forward equation of Eq.~\eqref{eq:1} equals that of a more simplified probability flow ordinary differential equation~\cite{song2021scorebased}. The implication of this observation is that both equations ultimately describe an identical statistical process, albeit in different form.

\begin{equation} \label{eq:2_2}
\begin{split}
    d \mathbf{x} &= \mathbf{f}(\mathbf{x}, t) \, dt - \frac12 \nabla \cdot [ \mathbf{g}(\mathbf{x}, t) \mathbf{g}(\mathbf{x}, t)^\top ] \, dt \\
    &- \frac12 \mathbf{g}(\mathbf{x}, t) \mathbf{g}(\mathbf{x}, t)^\top \nabla_\mathbf{x} \log p_t(\mathbf{x}) \, dt.
\end{split}
\end{equation}
Therefore, given a time-dependent model $\mathbf{s}_\theta(\mathbf{x}, t)$ that can approximate the score $\nabla_\mathbf{x} \log p_t(\mathbf{x})$, we can sample $\mathbf{x}_T \sim p_T$ and compute reverse steps according to Eq.~\eqref{eq:2_2} to obtain $\x_0 \sim p_0$ from the original data distribution.

\subsection{Grad-TTS}

In this section, we expound Grad-TTS, the score-based TTS model we use to demonstrate EdiTTS. A more in-depth discussion of its components, such as monotonic alignment search~\cite{kim2020glowtts}, can be found in the original paper~\cite{popov2021gradtts}.

The forward SDE of Grad-TTS is defined as

\begin{equation} \label{eqn:4}
    d \mathbf{x} = \frac12 \bm{\Sigma}^{-1} (\bm{\mu} - \mathbf{x}) \beta_t \, dt + \sqrt{\beta_t} \, d \mathbf{w},
\end{equation}
where $\bm{\mu}$ is the mean of a prior distribution, $\bm{\Sigma}$ is the covariance matrix, and $\beta_t$ is the noise schedule. The mean is estimated by a prior text encoder. In Grad-TTS, the covariance is assumed as an identity matrix to simplify computation; we keep this variable in the exposition for the sake of completeness.

Since the drift coefficient $\mathbf{f}(\cdot, t)$ is affine, the transition kernel $p_{0t}(\x_t \vert \x_0)$ is a Gaussian distribution whose mean and variance can be obtained in closed form. Let $\gamma_t = \text{exp} (- \smallint^t_0 \beta_s \, ds )$. The transition kernel can be written as

\begin{equation}
    p_{0t}(\mathbf{x}_t \vert \mathbf{x}_0) = \mathcal{N} \{(\mathbf{I} - \sqrt{\gamma_t})\bm{\mu} + \sqrt{\gamma_t} \mathbf{x}_0, \bm{\Sigma} (\mathbf{I} - \gamma_t) \}.
\end{equation}
In an infinite time horizon where $\lim_{t \to T} \gamma_t = 0$, the transition kernel simply converges to $\mathcal{N}(\bm{\mu}, \bm{\Sigma})$. In other words, the forward process guarantees that diffusion will yield samples from the prior distribution.

The reverse SDE of Grad-TTS can be derived using Eq.~\eqref{eq:2} and Eq.~\eqref{eqn:4}. 

\begin{equation} \label{eq:ode}
    d \mathbf{x} = \frac12 \left( \bm{\Sigma}^{-1} (\bm{\mu} - \mathbf{x}) - \nabla_\mathbf{x} \log p_t (\mathbf{x}) \right) \beta_t  \, dt.
\end{equation}
Therefore, given a score estimator $\mathbf{s}_\theta$ and an initial condition from the prior distribution $N(\bm{\mu}, \bm{\Sigma})$, we can compute Eq.~\eqref{eq:ode} to sample from the learned data distribution.

\subsection{SDEdit}

SDEdit~\cite{meng2021sdedit} is an SDE-based image editing framework that allows for stroke-based image synthesis. An image is first perturbed by coarse user edits, such as color strokes or rough hand-drawn sketches. The edited image is then diffused to obtain noisy priors from a Gaussian distribution. Throughout the reverse steps that follow, a binary mask is repeatedly applied to the intermediate representation at each step such that the non-edited region of the image retains its original form, whereas the edited portion is transformed by gradients computed by a score estimator that simulates the reverse SDE. When reversing the priors, the score estimator only relies on low frequency components to infer high frequency information according to its learned data distribution, thus synthesizing realistic images from the edited input. SDEdit can be applied to a vast array of tasks without any task-specific fine-tuning or a custom loss function.

\section{Proposed Method}
\label{sec:pagestyle}

Unlike score-based image models, which use the standard normal $\mathcal{N}(\mathbf{0}, \mathbf{I})$ as their prior, Grad-TTS uses spectrogram-like mean vectors predicted by a prior text encoder. Therefore, we cannot directly modify the input mel-spectrogram as in image editing, as modifying the input would result in a mismatch between the prior predicted from text and the input spectrogram. To circumvent this issue, we apply coarse edits to the encoded priors. Specifically, the input mel-spectrogram is first diffused without any user edits. Then, perturbations are directly applied on the sampled priors before the score estimator simulates the reverse SDE. These coarse perturbations approximate the effects of editing, similar to how hand-drawn color strokes are poor estimations of photo-realistic details. 

\subsection{Pitch Shift}

Pitch shifting is a task in which the pitch of an audio segment is modified according to user demand. Motivated by the simple definition of spectrograms as representations of frequencies over time, we directly shift the spectrogram-like latents $\bm{\tilde{\mu}}$ produced by the prior network along the frequency axis. An upward shift roughly corresponds to higher pitch, and vice versa. In practice, we pad the priors and convolve a weighted averaging kernel $\bm{K}_{ps}$ over $\bm{\tilde{\mu}}$ along the frequency axis while applying a binary mask $\bm{m}$ to ensure that non-editable regions of the spectrogram remain unchanged. $\x$ and  $\x_\text{edit}$ are each sampled from the original and perturbed prior distributions, respectively. At each reverse step, non-editable regions of $\x_\text{edit}$ generally follow $\x$, whereas editable regions are updated by score estimates computed from the perturbed prior. We employ a gradient softening mask $\bm{S}_g$ to better ensure that junctures between edited and non-edited segments are smooth and continuous. The pitch shift algorithm is schematically shown in Alg.~\ref{alg:editts-p}.

\begin{algorithm}[t]
\caption{Pitch shift algorithm for EdiTTS}\label{alg:editts-p}
\begin{algorithmic}
\Require $\bm{\tilde{\mu}}$ (time-aligned prior), $\bm{m}$ (mask for edited region), $\bm{K}_{ps}$ (pitch shift kernel), $\bm{S}_g$ (mask for softening gradient),  $\beta_{1}, ..., \beta_{T}$ (noise schedule), $T$ (total denoising steps)

\State $\bm{\tilde{\mu}_\text{edit}} \gets \bm{m} \odot (\bm{K}_{ps} * \bm{\tilde{\mu}}) + ( \mathbf{1} - \bm{m} ) \odot \bm{\tilde{\mu}} $
\State{$\bm{\epsilon} \sim \mathcal{N}(\bm{0}, \mathbf{I})$}
\State{$\mathbf{x} \gets \bm{\tilde{\mu}} + \bm{\epsilon}$}
\State{$\x_\text{edit} \gets \bm{\tilde{\mu}}_\text{edit} + \bm{\epsilon}$}

\For{$t = T, ..., 1$}
\State{$d \x_1 \gets \frac{\beta_{t}}{2T}(\bm{\tilde{\mu}} - \x - \mathbf{s_{\theta}}(\x, \bm{\tilde{\mu}}, t))$}
\State{$d \x_{2} \gets \frac{\beta_{t}}{2T}(\bm{\tilde{\mu}}_\text{edit} - \x_\text{edit} - \mathbf{s_{\theta}}(\x_\text{edit}, \bm{\tilde{\mu}}_\text{edit}, t))$}
\State{$\x \gets \x - d \x_{1}$}
\State{$\x_\text{edit} \gets \x_\text{edit} -((\mathbf{1}-\bm{S}_g) \odot d \x_{1} + \bm{S}_g \odot d \x_{2})$}
\EndFor

\State \Return{$\x_\text{edit}$}
\end{algorithmic}
\end{algorithm}

\subsection{Content Replacement}

Content replacement is a task in which an audio segment in one source is substituted with that from another. Here, we only consider a standard replacement task involving two audio files, but the approach can be extended to any number of audio sources. First, the two input waveforms are appropriately sliced to obtain $\bm{\tilde{\mu}^{a}_\text{src}, \tilde{\mu}^{b}_\text{trg}}$ and $\bm{\tilde{\mu}^{c}_\text{src}}$, where superscript $b$ denotes an editable region, and $a$ and $c$ are adjacent segments. We concatenate the three chunks using a soft mask $\bm{S}_c$ to produce a perturbed prior, $\bm{\tilde{\mu}_\text{edit}}$. Similar to pitch shifting, both $\x$ and $\x_\text{edit}$ are denoised simultaneously, with gradient softening masks applied throughout the reverse process. Alg.~\ref{alg:editts-c} delineates these steps in detail.

\begin{algorithm}[t]
\caption{Content replacement algorithm for EdiTTS}\label{alg:editts-c}
\begin{algorithmic}
\Require $\bm{\tilde{\mu}_\text{src}}$ (time-aligned source prior),  $\bm{\tilde{\mu}_\text{trg}}$ (time-aligned target prior), $\bm{m_\text{src}}$ (mask for edited region in $\tilde{\mu}_\text{src}$), $\bm{m_\text{trg}}$ (mask for extracted region in $\bm{\tilde{\mu}_\text{trg}}$), $\bm{S}_c$ (mask for softening concatenation), $\bm{S}_g$ (mask for softening gradient), $\beta_{1}, ..., \beta_{T}$ (noise schedule), $T$ (total denoising steps)

\State {$\bm{\tilde{\mu}^{a}_\text{src}}, \bm{\tilde{\mu}^{c}_\text{src}} \gets \mathtt{split\_source\_prior}(\bm{\tilde{\mu}_\text{src}}$, $\bm{m_\text{src})}$}
\State {$\bm{\tilde{\mu}^{b}_\text{trg}} \gets \mathtt{extract\_target\_chunk}(\bm{\tilde{\mu}_\text{trg}}$, $\bm{m_\text{trg}})$}
\State $\bm{\tilde{\mu}_\text{edit}} \gets \mathtt{soft\_concat}(\bm{\tilde{\mu}^{a}_\text{src}}, \bm{\tilde{\mu}^{b}_\text{trg}}, \bm{\tilde{\mu}^{c}_\text{src}}, \bm{S}_c)$
\State{$\bm{\epsilon} \sim \mathcal{N}(\mathbf{0}, \mathbf{I})$}
\State{$\x_\text{trg} \gets \bm{\tilde{\mu}_\text{trg}} + \bm{\epsilon}$}
\State{$\x_\text{edit} \gets \bm{\tilde{\mu}_\text{edit}} + \bm{\epsilon}$}

\For{$t = T, ..., 1$}
\State{$d \x_{1} \gets \frac{\beta_{t}}{2T}(\bm{\tilde{\mu}_\text{trg}} - \x_\text{trg} - \bm{s_{\theta}}(\x_\text{trg}, \bm{\tilde{\mu}_\text{trg}}, t))$}
\State{$d \x_{2} \gets \frac{\beta_{t}}{2T}(\bm{\tilde{\mu}_\text{edit}} - \x_\text{edit} - \bm{s_{\theta}}(\x_\text{edit}, \bm{\tilde{\mu}_\text{edit}}, t))$}
\State{$\x_\text{trg} \gets \x_\text{trg} - d \x_{1}$}
\State{$\x_\text{edit} \gets \x_\text{edit} -(\mathtt{slice\_and\_reshape}(d \x_{1}, \bm{S}_g) + (\mathbf{1} - \bm{S}_g) \odot d \x_{2})$}
\EndFor

\State \Return{$\x_\text{edit}$}
\end{algorithmic}
\end{algorithm}

Empirically, we found that soft concatenation reduced the occurrence of infelicitous mixing between the coda and onset of combined segments. Mispronunciation is attributed to the fact that the text encoder employs self-attention and convolutions, which facilitate token mixing on both local and global levels. As a result, a single frame not only contains information about the phoneme in its immediate index, but also bleeds hints about adjacent phonemes. Soft concatenation serves to weaken signals from non-immediate components near frame junctures. 

\section{Experiments}

\subsection{Setup}

EdiTTS was evaluated on the LJ speech dataset~\cite{ito2017lj}. We conducted two experiments to separately assess pitch shifting and content editing algorithms, employing HiFi-GAN~\cite{kong2020hifi} as the vocoder. We also used the official open-sourced checkpoint from Grad-TTS~\cite{popov2021gradtts}. The number of reverse steps $T$ was set to 1000 during inference. The value of the softening gradient mask $\bm{S}_g$ corresponding to editable regions of the audio was set to 1, whereas non-editable segments were masked according to $\sum_{k=i}^{16}{2^{16-k}} / \sum_{k=0}^{16}{2^{k}}$, where $i \in [1, 16]$ is the positional difference between the current index and the nearest editable region. The remainder of $\bm{S}_g$ was set to 0. Similarly, non-editable regions of the softening concatenation mask $\bm{S}_c$ were set to $0.1 \times (10-j) $, where $j \in [1, 9]$ is the positional difference as defined above. The remainder of $\bm{S}_c$ was set to zero. For upward pitch shifting, we set $\bm{K}_{ps}$ to (0.2, 0.2, 0.6, 0, 0); downward kernel, (0, 0, 0.6, 0.2, 0.2).

\label{eval}
\subsection{Pitch Shift}

We compared EdiTTS with two existing baselines: FastPitch~\cite{lancucki2021fastpitch} and WORLD~\cite{masanori2016world}. Additionally, we considered an ablated formulation of EdiTTS, denoted as ``Mel shift" in Table~\ref{table:mos}, in which only coarse mel-spectrogram shifting was applied on the output of Grad-TTS without denoising reversal. We conducted a mean opinion score (MOS) test with 18 participants through Amazon Mechanical Turk, a crowd-sourcing platform. 15 sentences were randomly chosen from the LJ speech dataset test split. For each sentence, we randomly sampled an editable target region, which was used to generate pitch-unchanged, pitch-up, and pitch-down utterances. Given a pair of pitch-unchanged and adjusted samples, participants were asked to provide a holistic score on a five-point scale, based on (1) how natural the pitch-adjusted sample is overall, and (2) how discernible the edited pitch is compared to the reference sample. 

\begin{table}
  \caption{Mean Opinion Scores with 95\% confidence interval}
  \label{table:mos}
  \centering
  \setlength\tabcolsep{2.3pt}
  \begin{tabular}{c|c|c|c}
    \toprule
    Method & Pitch Up & Pitch Down & Overall \\
    \midrule
    WORLD~\cite{masanori2016world} & $2.79 \pm 0.12$ & $2.98 \pm 0.13$ & $2.88 \pm 0.09$ \\
    FastPitch~\cite{lancucki2021fastpitch} & $3.77 \pm 0.10$ & $3.78 \pm 0.10$ & $3.78 \pm 0.07$ \\
    Mel shift & $3.75 \pm 0.10$  &  $3.69 \pm 0.09$ & $3.73 \pm 0.07$  \\
    EdiTTS (Ours) & $\mathbf{4.04 \pm 0.09}$ & $\mathbf{3.86 \pm 0.10}$ & $\mathbf{3.95 \pm 0.07}$  \\
    \bottomrule
  \end{tabular}
\end{table}

As shown in Table \ref{table:mos}, EdiTTS outperforms WORLD and FastPitch. In particular, EdiTTS's pitch-up score lies beyond the confidence interval in comparison to the baseline methods, suggesting statistical significance. The ablated baseline performed on par with FastPitch. We hypothesize that this is due to the relatively unnatural prosodic patterns exhibited in FastPitch samples. On the other hand, the ablated baseline leverages well-formed mel-spectrograms generated by Grad-TTS, albeit with some artifacts introduced by kernel convolution. The score discrepancy between EdiTTS and the mel shift baseline demonstrates that post-synthesis denoising edit is an integral component of EdiTTS, and that spectral perturbation alone is insufficient to achieve satisfactory output quality.

\begin{figure}[t]
  \centering
  \includegraphics[width=0.95 \linewidth]{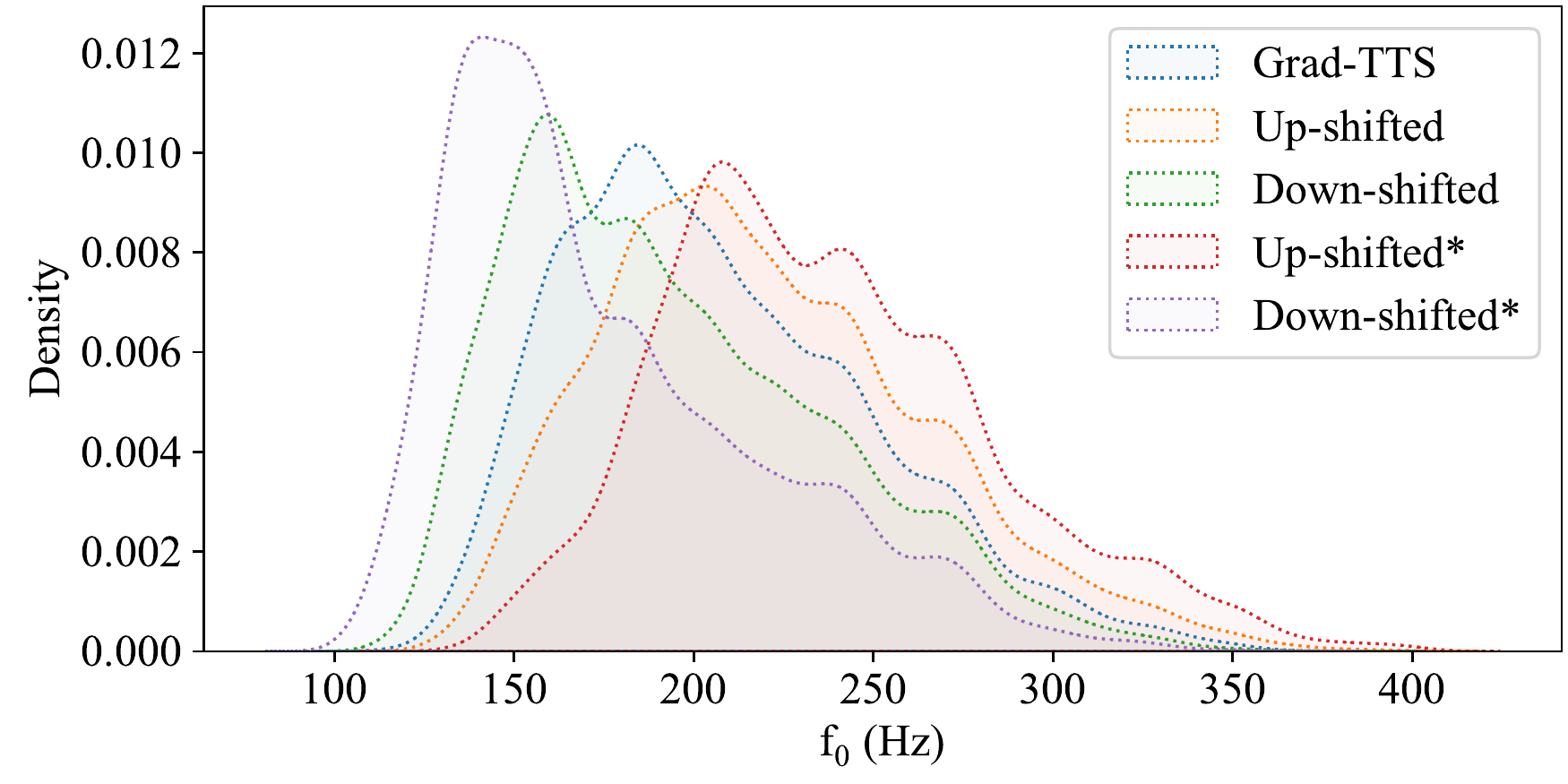}
  \caption{$f_0$ statistics of pitch-altered samples}
  \label{fig:histogram}
\end{figure}

To further investigate the effectiveness of the pitch shifts created by EdiTTS, we extracted $f_0$ values of the edited regions using CREPE~\cite{kim2018crepe}. We scrutinized the relationship between the pitch shift kernel and the magnitude of the resulting pitch shift. To this end, EdiTTS was configured with two different kernels: the default $\bm{K}_{ps}$ at (0.2, 0.2, 0.6, 0, 0), and the more aggressive mode at (0.4, 0.4, 0.2, 0, 0), denoted in Fig.~\ref{fig:histogram} with an asterisk (*). The density plot shows a modal progression from aggressive-down, default-down, Grad-TTS, default-up, and aggressive-up configurations. In particular, we found that the pitch variation of the modified setup is approximately double that of the default mode, suggesting that the configuration of the pitch shift kernel offers direct control over the magnitude of the pitch shift in the edited sample. 

\begin{figure}[t]
  \centering
  \includegraphics[width=0.95 \linewidth]{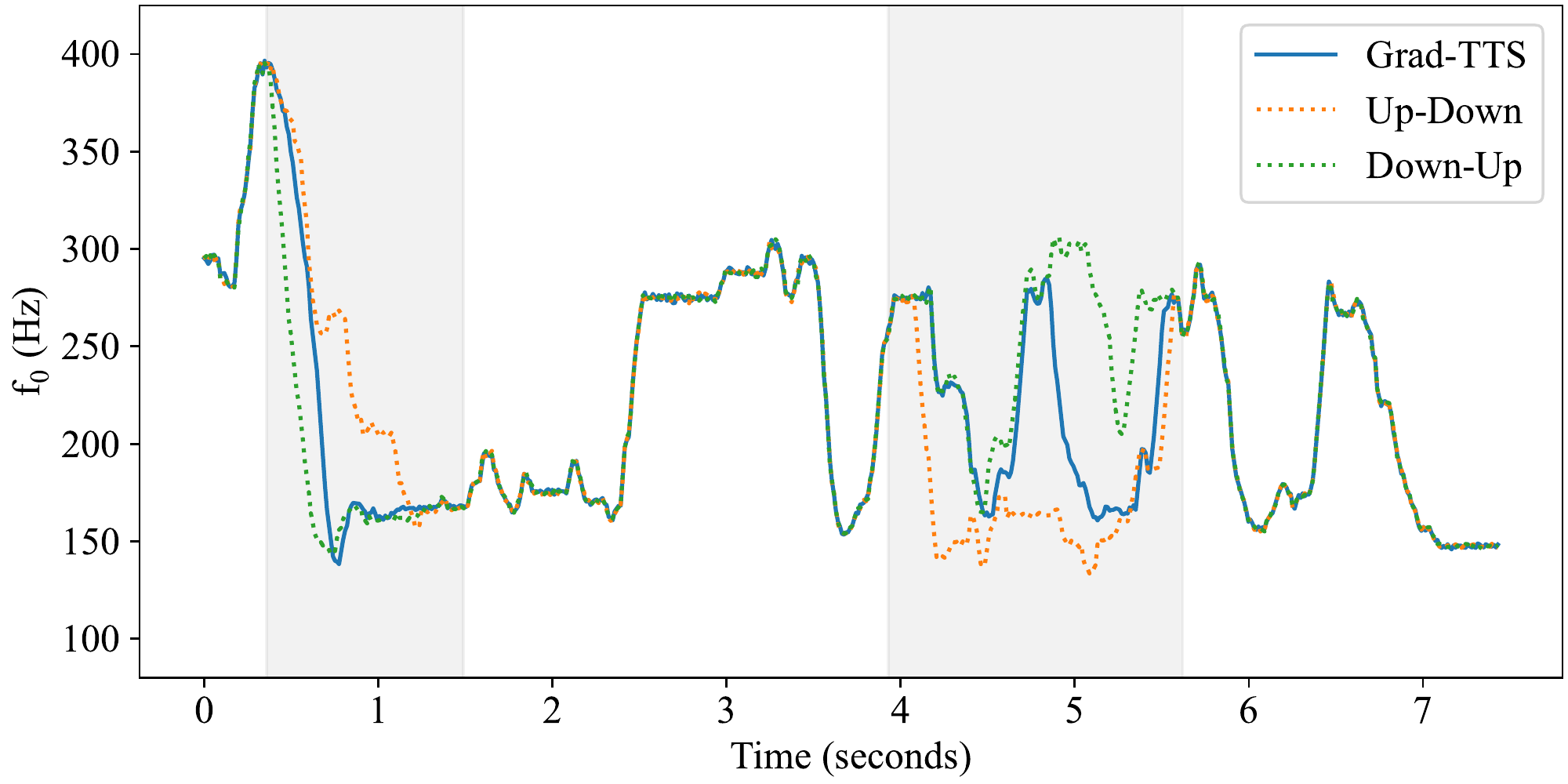}
  \caption{$f_0$ plot of up-down and down-up samples}
  \label{fig:sample}
\end{figure}

We performed additional sample-level qualitative experiments for closer examination of pitch shift patterns. Fig.~\ref{fig:sample} illustrates a plot of an up-down and down-up pitch alteration sequence with two discrete editable regions on a single randomly chosen sentence. Shaded regions represent edited segments. We were able to empirically verify that EdiTTS can apply multiple edits to a single sample, and that doing so does not affect other non-editable regions unspecified by the user. 

\begin{figure}[t]
  \centering
  \includegraphics[width=0.95 \linewidth]{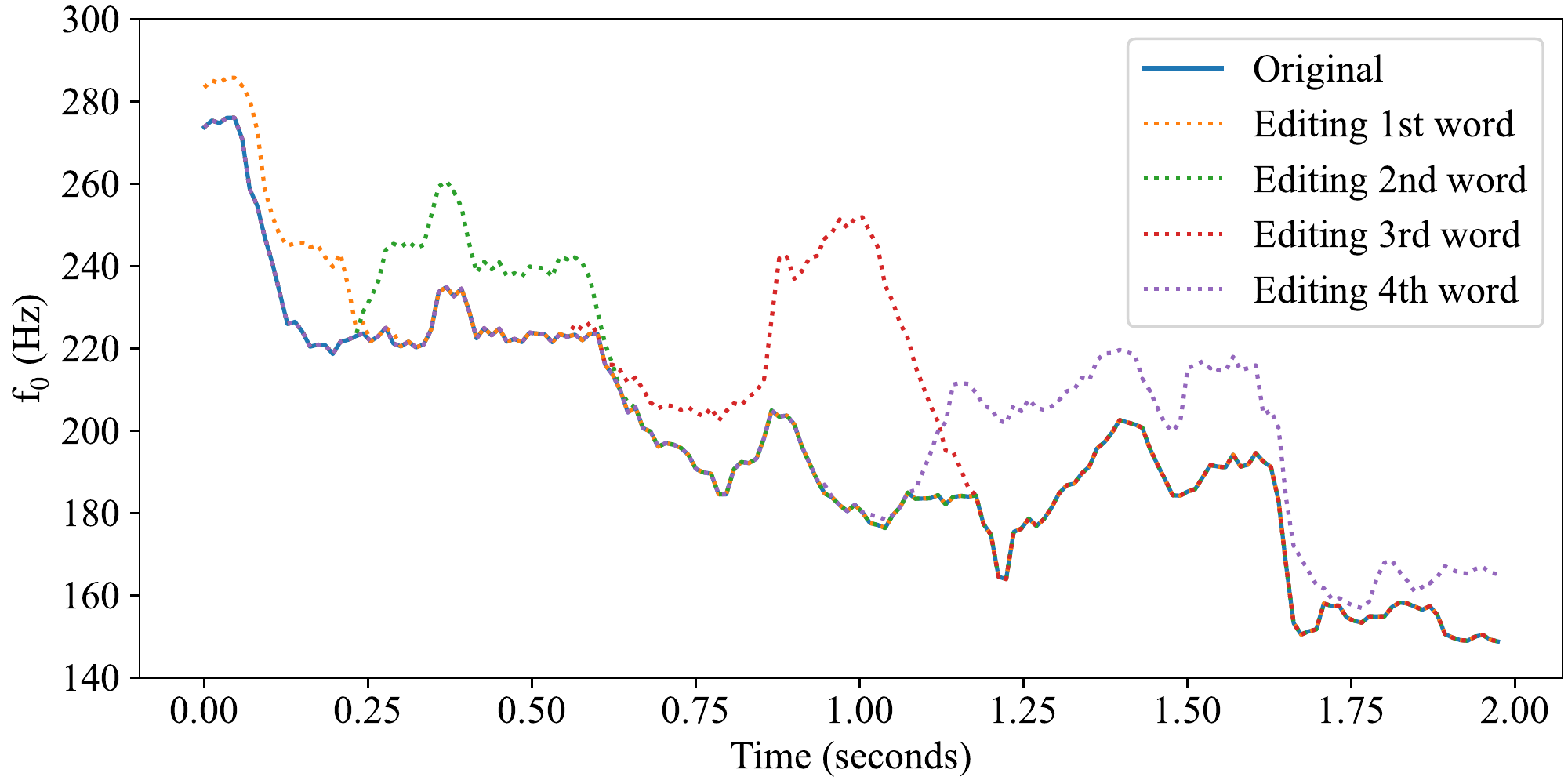}
  \caption{$f_0$ plot of word-level pitch altered samples}
  \label{fig:one}
\end{figure}

Another topic of interest was the granularity of edits afforded by EdiTTS. A highly realistic application of speech editing is the modification of $f_0$ to remove or create accentuation in a recording. We applied pitch-up shifts on each word of a LJ Speech test sentence, \textit{`from August nineteen sixty-two'} and report the result in Fig~\ref{fig:one}. We observed that the pitch of each target word increased properly while leaving the remaining parts unchanged. In particular, a non-trivial pitch shift was achieved in the third word, creating discernible accentuation. 

\subsection{Content Replacement}

For content edit evaluation, we compared EdiTTS against Grad-TTS~\cite{popov2021gradtts} and naive spectrogram concatenation, in which spectrogram frames from disparate audio sources were simply joined together to emulate the effects of content substitution.

We prepared 80 test sentences from the LJ speech dataset by replacing one random word with its synonym using the \texttt{nltk} library~\cite{nltk}. The resulting sentences were manually reviewed by a native speaker to ensure that they read naturally. For each sentence, we synthesized two audio samples: one through Grad-TTS using the synonym-replaced script, and another through EdiTTS, where the selected word in the original audio sample was replaced with a synthesized utterance of its synonym. We used the Google Cloud speech-to-text API~\cite{gcloud} to examine the word error rate (WER) and the character error rate (CER) of content edits performed by EdiTTS.

\begin{table}[t]
  \caption{Word and character error rates}
  \label{table:er}
  \centering
  \begin{tabular}{c|c|c}
    \toprule
    Method & WER (\%) & CER (\%) \\
    \midrule
    Mel concat & 11.6129 & 4.4241 \\
    Grad-TTS~\cite{popov2021gradtts} & 8.2258  &  2.8215 \\
    EdiTTS (Ours) & 9.1129 & 3.2187  \\
    \bottomrule
  \end{tabular}
\end{table}

As shown in Table \ref{table:er}, EdiTTS substantially outperforms the naive mel concatenation baseline and marginally lags behind Grad-TTS. Note that the error rates achieved by Grad-TTS can be viewed as upper bounds for our method since EdiTTS itself employs Grad-TTS for spectrogram synthesis. We observe a considerable gap between the error rates given by the baseline and Grad-TTS. Concatenating mel-spectrograms extracted from different utterances introduces prosodic discontinuities and audible artifacts, which adversely affect the error rate. In contrast, EdiTTS processes the concatenated input via denoising reversal, which smoothens such irregularities. The discrepancy between the error rates of naive spectrogram concatenation and EdiTTS indicates that score-based denoising of perturbed priors is integral to improving the quality of the final output.

\section{Limitations}
\label{limitations}

We discuss some limitations of EdiTTS and ways to improve it.

\textbf{Speed.} Generating samples from diffusion models is slow due to the iterative nature of their sampling process. Because EdiTTS relies on the reverse process of the score-based backbone, the statement also applies. Improving the sampling speed of diffusion models is an ongoing area of research~\cite{nichol2021improved}, and we fully expect these developments to give rise to better diffusion-based TTS and speech editing methodologies.

\textbf{Specificity.} 
Although EdiTTS enables different degrees of pitch shifts depending on the pitch shift kernel, the exact amount of shift cannot be specified. This contrasts with FastPitch or WORLD, which explicitly accepts $f_0$ values to generate samples that follow the input condition. One possible step towards improvement in this direction would be to apply different kernels for each micro-segment within the editable region instead of fixing a single $\bm{K}_{ps}$ for all frames.

\textbf{Alignment.} EdiTTS relies on text-to-speech alignments found by Grad-TTS to modify Gaussian latents that correspond to user-selected text. While the error rates reported in Table~\ref{table:er} suggest that misalignments rarely occur, EdiTTS could edit irrelevant portions of the audio if inaccurate alignments are found. We believe ongoing advancements in robust TTS alignments~\cite{badlani2021tts} will help improve not only EdiTTS, but also non-autoregressive TTS pipelines in general.

\section{Conclusion}
\label{conclusion}

We propose EdiTTS, a score-based speech editing methodology for fine-grained pitch and content editing. EdiTTS applies coarse perturbations in the diffused prior space to induce desired edits from the model during denoising reversal. A marked advantage of EdiTTS is its ease of applicability: we leverage the TTS backbone as-is without any additional training or auxiliary module for each editing task. Through listening tests and error rate analyses, we demonstrate that EdiTTS achieves competitive edge over simple mel-spectrogram engineering, as well as performing on par with the original Grad-TTS and outperforming existing baselines on standard speech editing tasks.

\section{Acknowledgement}
This research was supported by the Korean Ministry of Culture, Sports and Tourism and the Korea Creative Content Agency (project number: R2021050007).

\bibliographystyle{IEEEtran}

\bibliography{refs}

\end{document}